\documentstyle[aps,multicol,prl]{revtex}

\begin{document}

\bibliographystyle{prsty}
\draft

\title{The long-wavelength behaviour of the exchange-correlation kernel in the
Kohn-Sham theory of periodic systems}
\author{Ph. Ghosez$^1$, X. Gonze$^1$ and R. W. Godby$^2$}
\address{$^1$Unit\'e de Physico-Chimie et de Physique des Mat\'eriaux,
Universit\'e Catholique de Louvain,\\
1 Place Croix du Sud, B-1348 Louvain-la-Neuve, Belgium}
\address{$^2$Department of Physics, University of York, Heslington,
York YO1 5DD, U. K.}
\date{\today}
\maketitle
\begin{abstract}
The polarization-dependence of the exchange-correlation (XC) energy
functional of periodic insulators within Kohn-Sham (KS) density-functional
theory
requires a ${\cal O} (1/q^2)$ divergence in the XC kernel
for small vectors {\bf q}. This
behaviour, exemplified for a one-dimensional model semiconductor,
is also observed when an insulator happens to be
described as a KS metal, or vice-versa. Although
it can occur in the exchange-only kernel,
it is not found in the usual local, semi-local or even non-local
approximations to KS theory. We also
show that the test-charge and electronic definitions of the macroscopic
dielectric
constant differ from one another in exact KS theory,
but are equivalent in the above-mentioned approximations.

PACS numbers: 71.10.+x, 77.22.Ej
\end{abstract}
\begin{multicols}{2}[]
\setcounter{page}{1}

\section{introduction}

We have recently reexamined \cite{Gonze95,Gonze97,OEP}
the exchange-correlation potential, $V_{\rm xc}({\bf r})$, that
represents electronic exchange and correlation in the Kohn-Sham formulation
of density-functional theory (DFT), in the case of macroscopic bodies.
Our analysis focused on periodic solids with a
uniform macroscopic polarization: a crystal subjected to a homogeneous external
electric field, or a polar crystal.  We showed that the exact $V_{\rm xc}$
inside the crystal is required to have an ultra-non-local dependence
on the surface electron
density, or, equivalently, on the macroscopic polarization.  In the present
paper,
we will
show that the exchange-correlation kernel, $K_{\rm xc}$, is a powerful
vehicle for analyzing the presence of this dependence in the common
approximations
for exchange and correlation.
The required wavevector-dependence of $K_{\rm xc}$ is illustrated in the
exact DFT of
a model system.  Links with the polarizability in insulators and metals
will also be established.

We first define some useful quantities and notation.
Within the Kohn-Sham (KS) formulation of Density Functional Theory \cite{DFT},
a system of interacting electrons in its ground state,
placed in an external potential, $V_{\rm ext}({\bf r})$, is
mapped onto a fictitious system of independent particles in an
effective potential
\begin{equation}
V_{\rm KS}({\bf r})=
V_{\rm ext}({\bf r})+V_{\rm H}({\bf r})+V_{\rm xc}({\bf r}),
\label{Eq.1}
\end{equation}
under the requirement that both generate the
same density $n({\bf r})$.
The combined external $V_{\rm ext}({\bf r})$ and
Hartree $V_{\rm H}({\bf r})$ potentials
give the electrostatic potential $V_{\rm TC}({\bf r})$ felt by a classical
test charge (TC),
while the exchange-correlation (XC) term, $V_{\rm xc}({\bf r})$,
that subsumes all the additional many-body effects, acts only on KS electrons.

The XC potential at point {\bf r} is the first derivative
of the XC energy with respect to the density at this point,
\begin{equation}
V_{\rm xc}({\bf r}) = \frac{\delta E_{\rm xc}}{\delta n({\bf r})}.
\label{Eq.Vxc}
\end{equation}
The second derivative of this XC energy with respect to the densityis
called the
XC kernel:
\begin{equation}
K_{\rm xc}({\bf r},{\bf r'}) =
\frac{\delta^2 E_{\rm xc}}{\delta n({\bf r}) \delta n({\bf r'})}.
\label{Eq.Kxc}
\end{equation}
The explicit
forms of $E_{\rm xc}$, $V_{\rm xc}$ or $K_{\rm xc}$ as functionals of the
density
are unfortunately unknown and calculations are usually
performed within the local-density approximation (LDA),
within semi-local approximations such as the generalized gradient
approximations (GGA)~\cite{Khein95}, or even within non-local approximations
such as the weighted-density approximation (WDA)~\cite{WDA}.

In the same manner as $V_{\rm xc}$ constitutes a sizeable contribution to the KS
effective potential, $K_{\rm xc}$ plays an important role
in the investigation of the responses of the KS system
to static external pertubations.
The density response of the interacting electron system to an external
perturbation is described by the
polarizability matrix $\chi$:
\begin{equation}
\delta n({\bf r}) =
\int
\chi({\bf r},{\bf r'}) \;
\delta V_{\rm ext}({\bf r'}) d{\bf r'}.
\end{equation}
The so-called ``proper part'' of the density response,
$\pi$~\cite{Pick70}, relates
the change of density to the change of the test-charge potential:
\begin{equation}
\delta n({\bf r}) =
\int
\pi({\bf r},{\bf r'}) \;
\delta V_{\rm TC}({\bf r'}) d{\bf r'}.
\end{equation}
{}From the relation between TC and external potentials,
these matrices are related by the following expression
(making use of matrix notations):
\begin{equation}
\pi^{-1} =
\chi^{-1} + V_C,
\label{Eq.pi}
\end{equation}
where $V_{\rm C}$ stays for the Coulomb interaction.
For periodic solids, the various quantities
are most conveniently represented in reciprocal space, where
$\chi({\bf q}+{\bf G},{\bf q}+{\bf G'})$, for example, may be thought
of as a matrix in the reciprocal-space
lattice vectors ${\bf G}$ and ${\bf G'}$, for each wavevector ${\bf q}$, with
the element ${\bf G}={\bf G'}={\bf 0}$ known as the head, the other elements
in that row and column as the wings, and the rest of the matrix as the body
\cite{HeadWingBody}.

For the KS system, an independent-particle polarizability
matrix $\chi_o$ is usually
introduced, relating the change of density to that of the effective KS
potential:
\begin{equation}
\delta n({\bf r}) =
\int
\chi_o({\bf r},{\bf r'}) \;
\delta V_{\rm KS}({\bf r'}) d{\bf r'}.
\end{equation}
Using Eqs.~(\ref{Eq.1})~and~(\ref{Eq.pi}),
this independent-particle polarizability matrix is easily linked with the
interacting-particle polarizability and its proper part:
\begin{eqnarray}
\chi_o^{-1} && =
\chi^{-1} + V_C + K_{\rm xc}
\nonumber
\\
&& = \pi^{-1} + K_{\rm xc}.
\label{X-Xo}
\end{eqnarray}
Following the Adler and Wiser sum-over-state technique~\cite{Adler62+},
$\chi_o$ can be directly computed from the KS wavefunctions.
As shown by Eq.~(\ref{X-Xo}), the knowledge of $K_{\rm xc}$
is crucial in deducing $\chi$ or $\pi$ from $\chi_o$. This key role was
recently recalled by Dal Corso, Baroni and Resta~\cite{DalCorso94}.

In connection with the KS band-gap problem
it was shown by Godby and Needs~\cite{Godby89}, within certain approximations,
that the ground state of a periodic insulator
is sometimes described as a metal in KS theory. This is particularly
striking in that the long-wavelength behaviour of the polarizability matrix
$\chi$ for an insulator is qualitatively different from that of a metal, while
$\chi$ is a ground-state quantity and should be correctly obtained within DFT.
Godby and Needs emphasized that a highly nonanalytic $K_{\rm xc}$
could allow such a phenomenon.

Recently, we proposed that the exact XC energy functional
is polarization-dependent in the case of periodic insulating solids
submitted to a homogeneous electric field~\cite{Gonze95}.
After exploring in some detail the consequences of this
finding, Aulbur, J\"onsson and
Wilkins~\cite{Aulbur97} showed the presence of a sizeable contribution
of the polarization-dependence of $E_{\rm xc}$ to the
linear and non-linear optical response of real materials.
For polar solids, careful treatment of the polarization
is mandatory, even if no homogeneous electric field is present~\cite{Gonze97}.
Resta~\cite{Resta96} tried to address
the origin of these effects: he proposed that the
polarization dependence should arise from the Coulomb coupling between the
electron, in the bulk, and part of its correlation hole delocalized at the
surface. At variance, Martin and Ortiz~\cite{Martin96}, pointed out that the
shape of the XC hole may already depend on the surface
charge, and we demonstrated~\cite{OEP} that a polarization-dependence is
expected at the purely exchange level (without correlation).
In another paper,
Martin and Ortiz~\cite{Martin97} placed the new density-polarization
functional theory in the perspective of important works of the seventies,
and presented an alternative formulation of it.

In Section II of the present paper, we link the polarization-dependence
of the XC energy to a
${\cal O} (1/q^2)$ divergence of $K_{\rm xc}$ in the limit of $q \rightarrow 0$,
briefly sketched in
Ref.~\cite{Gonze96}, and illustrate this behaviour in the case of a
one-dimensional
semiconductor. An analysis of related points was given in Ref.~\cite{Aulbur97}.
This requirement of the exact KS theory is not fulfilled in the usual
approximations such as the LDA, the GGA, and the WDA, as shown in Section III.
In Section IV, we observe that a ${\cal O} (1/q^2)$ behaviour
of the exact XC kernel is able to resolve the
``true insulator - KS metal'' paradox. A unified treatment of
$K_{\rm xc}$ for insulators and metals is given in Section V, together with
an analysis of the exchange-only kernel.
We will finally discuss (Section VI) the consequences of
these facts for the different
definitions of the macroscopic dielectric constant.

\section{ Divergence of $K_{\rm xc}$}

Working on periodic insulators first,
we adopt the same perturbative approach (long-wave
method) as in Ref.~\cite{Gonze95}, and briefly recall one of its central
result.
The change of external potential $\delta
V_{\rm ext}$ produced by an infinitesimal homogeneous electric field $\delta
{\cal E}_{\rm ext}$ is (written in one dimension for brevity):
\begin{eqnarray}
\delta V_{\rm ext} (r)
&=&
\lim_{q \rightarrow 0} \; \delta V_{\rm ext} (q) \; (e^{iqr}- e^{-iqr})
\nonumber \\
&=&
\lim_{q \rightarrow 0} \; \frac{\delta {\cal E}_{\rm ext}}{q} \; \sin(qr)
\end{eqnarray}
In reponse to this perturbation, the system will develop a change of density
$\delta n(r)$. Owing to local field effects (Umklapp processes),
it may contain contributions at
different $(q+G)$ vectors (where $G$ belongs to the
reciprocal lattice). Within linear response, the long-wave part of
$\delta n$ takes the form:
\begin{equation}
\delta n (r)
=
 - \; \lim_{q \rightarrow 0} \; q \; \delta {\cal P} \; \sin(qr)
\end{equation}
where $\delta {\cal P}$ is the change of polarization for $q=0$~\cite{sign}.

Generally, the self-consistent screening potential will also contain
long-wave and more rapidly varying terms. Its long-wave part will
include an Hartree contribution, corresponding to the screening
of the applied field due to the Coulomb interaction. In Ref.~\cite{Gonze95},
we demonstrated that the polarization dependence of
$E_{\rm xc}$ will manifest itself
through a homogeneous XC electric field
$\delta {\cal E}_{\rm xc}$, so that for $q \rightarrow 0$:
\begin{equation}
\delta V_{\rm xc} (q) =
\frac{\delta {\cal E}_{\rm xc}}{2 i q} \; .
\end{equation}

{}From Eqs.~(\ref{Eq.Vxc}) and (\ref{Eq.Kxc}),
the XC kernel and potential are related by $\delta
V_{\rm xc} = K_{\rm xc} \,  \delta n$. Isolating in this equation
the long-wave terms from the other contributions,
we obtain, in a generalized matrix notation (in which $G$ stands
for {\it all} non-zero vectors of the reciprocal lattice):
\begin{eqnarray}
\left(
\begin{array}{c}
\frac{1}{q} \; \frac{\delta {\cal E}_{\rm xc}}{2i} \\
\delta V_{{\rm xc},{q+G}}
\end{array}
\right)
 =
K_{\rm xc}
\left(
\begin{array}{c}
- { q} \; \frac{\delta {\cal P}}{2i} \\
\delta n_{q+G}
\end{array}
\right) \; .
\end{eqnarray}
In order for the change in exchange-correlation field $\delta {\cal E}_{\rm xc}$
to be finite when a finite change of polarization $\delta {\cal P}$ takes
place, the
head of the exact exchange-correlation kernel matrix, $K_{\rm xc}(q,q)$, must
exhibit a ${\cal O} (1/q^2)$ divergence in the limit of $q \rightarrow 0$
~\cite{Gonze96,Aulbur97}.


The previous results can be illustrated in the case of a simple
model one-dimensional semiconductor,
already used in Ref.~\cite{Gonze95}, and for
which $K_{\rm xc}$ can be computed exactly (within the model). In this model,
the sum of the external and Hartree potential is taken to be:
\begin{equation}
V_{\rm ext}(x)+V_{\rm H}(x) = V_o \cos (2 \pi x/a),
\end{equation}
where $a$ is the unit cell length. Moreover, a simple non-local self-energy
operator
is present, with the aim of mimicking the relevant many-body effects. It has
the same non-local form and same parameters as in Refs.~\cite{Godby94,Gonze95}:
\begin{equation}
\Sigma(r,r',\omega) =
\frac{f(x)+f(x')}{2} \; g(|x-x'|)
\end{equation}
 where $f(x)= F_o \; [1 - \cos (2 \pi x/a)]$ is a negative function that
has the cell
periodicity and $g(y)$ is a normalized gaussian of width $w$. We construct
an {\it exact} KS theory for this model system
by determining the local potential $V_{\rm KS}$, which,
when filled with non-interacting electrons, reproduces the density
obtained when including the self-energy operator~\cite{Godby94,Gonze95}.

The proper part $\pi$ of the polarizability of the model system can be obtained
using the Adler and Wiser~\cite{Adler62+} sum-over-state technique applied
to the eigenfunctions and eigenvalues of the reference Hamiltonian (that
includes $\Sigma$, which is taken to be independent of
changes in the external potential).

{}From our KS wavefunctions, that reproduce the same density, it is also
possible to compute the KS independent-particle polarizability, $\chi_o$,
from the Adler and Wiser technique. The relationship
between $\pi$ and  $\chi_o$ is given by Eq.~(\ref{X-Xo}), so that we obtain~:
\begin{equation}
K_{\rm xc} =   \chi_o^{-1}  - \pi^{-1}.
\label{Eq.11}
\end{equation}

In Fig.~\ref{Fig.Kxc} we have
plotted the diagonal part of the computed
$K_{\rm xc}$. The calculation was performed on a 80-unit-cell,
which guarantees a convergence better than 0.7\%.
We observe a divergence in the limit of $q \rightarrow 0$ . The inset
exhibits its expected ${\cal O} (1/q^2)$ character.

We note that the divergence of $K_{\rm xc}$ has been
obtained without including long-range correlation
effects in our model semiconductor~\cite{Comment2}.
Inclusion of such effects would simply modify the
coefficient of the $K_{\rm xc}(q,q)$ divergence.

\section{Approximate functionals}

In the previous Section, we have linked the polarization-dependence of the XC
energy with an ${\cal O} (1/q^2)$ divergence of the XC kernel,
and given an example of this behaviour.
The link with a third concept, the ultra-non-local sensitivity of the
exchange-correlation functional~\cite{Godby94} was emphasized in
Refs.~\cite{Gonze95,Gonze97}: a
change in surface charge may have an influence on the
bulk XC potential, independently of the distance between the
point in the bulk and the surface.
Any approximate XC functional can now be analyzed
in the light of these characteristics of the exact functional. Although they
have the same physical origin, each of them provides a different point of view.

As regards the ultra-non-locality requirement, the behaviour of LDA and GGA
is clear: the corresponding exchange-correlation potentials at any point
do {\it not} depend on the density outside the immediate
neighborhood.
The following analysis of the small wavevector
dependence of the exchange-correlation kernel provides a more refined picture of
the violation of this requirement.

To start with, we consider the XC kernel in the LDA,
\begin{equation}
K_{\rm xc}^{\rm LDA}({\bf r, r'}) =
\left. \frac{ \delta V^{\rm LDA}_{\rm xc} }{ \delta n}
\right|_{\bf r} \; . \; \delta ({\bf r-r'}).
\end{equation}
The Fourier transform
of this kernel, diagonal in real space, is such that
$K_{\rm xc}^{\rm LDA}({\bf q,q})$ is {\it independent}
of {\bf q}~\cite{Hybertsen87}.

The gradient-corrected XC energy has the form
\begin{equation}
E_{\rm xc}^{\rm GGA}[n]=\int e_{\rm xc}
  [
      n({\bf r}), {\bf \nabla} n({\bf r})
  ]
     d{\bf r},
\end{equation}
with the corresponding potential~:
\begin{equation}
V_{\rm xc}^{\rm GGA}({\bf r}) =
{\left.
\frac{ \partial e_{\rm xc} }{ \partial n }
\right|
}_
{\bf r}
- {\bf \nabla} \cdot
{\left.
\frac{ \partial e_{\rm xc} }{ \partial  {\bf \nabla}  n }
\right|
}_{\bf r}
\end{equation}
The relationship between the long wavelength
GGA-XC potential and the long wavelength density
is therefore governed by the following kernel
(compare with Eqs.(15) and (22) of Ref.~\cite{DalCorso94}):
\begin{eqnarray}
K_{\rm xc}^{\rm GGA}({\bf q},{\bf q})=
\frac{1}{\Omega} \, \,
\biggl[
\,\,\,\,\,\,
&&
\int_{\Omega}
{\left.
    \frac{ \partial^2 e_{\rm xc} }{ \partial n^2 }
   \right|
 }_{\bf r}
d{\bf r}
\nonumber \\
-
2 i \sum_\alpha q_\alpha
\;
&&
\int_{\Omega}
{\left.
    \frac{ \partial^2 e_{\rm xc} }{ (\partial n) (\partial (\partial_\alpha
n)) }
   \right|
 }_{\bf r}
d{\bf r}
\nonumber \\
- \sum_{\alpha \beta}
q_\alpha q_\beta
\;
&&
\int_{\Omega}
{\left.
    \frac{ \partial^2 e_{\rm xc} }{ (\partial (\partial_\alpha n))
                            (\partial (\partial_\beta n)) }
   \right|
 }_{\bf r}
d{\bf r}
\,\,
\biggr]
\label{EqGGA}
\end{eqnarray}
The first term of the right member of this equation
is the Fourier transform of the term appearing in the LDA.
Eq.~(\ref{EqGGA}) makes clear that
the gradient corrections produces terms with {\it positive} powers
of $q$, but not the required ${\cal O}(1/q^2)$ divergence.

Contrary to what is
expected by Mazin and Cohen~\cite{Mazin97}, the GGA has therefore no apparent
ability to improve upon the LDA behaviour in this respect. Mazin and Cohen were
also expecting the weighted-density approximation (WDA), in which the XC
functional
is truly non-local, to improve upon the LDA behaviour.
We now analyze this case.

The exchange-correlation energy is given exactly by
the adiabatic connection formula,
in terms of the coupling-constant averaged
pair-correlation hole, $G({\bf r},{\bf r'};[n])$:
\begin{eqnarray}
E_{\rm xc}[n]= \frac{1}{2} \int \!\!\! \int
	n({\bf r})
 \frac{ G({\bf r},{\bf r'};[n]) } { |{\bf r} - {\bf r'}| }
 n({\bf r'})
 d{\bf r} d{\bf r'}
\end{eqnarray}
This exact expression was the starting point
of the analysis of the XC hole by Resta~\cite{Resta96}.
In the WDA~\cite{WDA}, the exact pair-correlation function at each point
${\bf r}$
is replaced by the pair-correlation
evaluated for a homogeneous electron gas with``weighted'' density,
 $G^{\rm hom}({\bf r},{\bf r'};\bar n({\bf r};[n]))$, where
$\bar n({\bf r};[n])$ is determined from the density in the whole space
so as to
satisfy the requirement that the exchange-correlation
hole integrates to $-1$.
The corresponding WDA-XC potential is made of three terms
\begin{eqnarray}
V_{\rm xc}^{\rm WDA}({\bf r}_0)=
\frac{1}{2} \int
 \frac{ G^{\rm hom}({\bf r}_0,{\bf r'};\bar n({\bf r};[n])) } { |{\bf r}_0 -
{\bf r'}| }
 n({\bf r'})
 d{\bf r'}
\nonumber
\\
+
\frac{1}{2} \int n({\bf r})
 \frac{ G^{\rm hom}({\bf r},{\bf r}_0;\bar n({\bf r};[n])) } { |{\bf r} - {\bf
r}_0| }
 d{\bf r}
\nonumber
\\
+
\frac{1}{2} \int \!\!\!\! \int	
\frac{ n({\bf r}) n({\bf r'}) } { |{\bf r} - {\bf r'}| }
\frac{ \delta G^{\rm hom}({\bf r},{\bf r'};\bar n({\bf r};[n])) } { \delta
n({\bf r}_0) }
 d{\bf r} d{\bf r'}.
\label{Eq.WDA}
\end{eqnarray}
If the WDA-XC energy functional were ultra-non-local, arbitrarily remote
surface charges would influence the XC potential, so as to induce the
possibility
of a linear XC potential in the bulk. The first or second terms of
Eq.~(\ref{Eq.WDA}) would present such a behaviour if the
homogeneous gas pair-correlation
function $G^{\rm hom}$ were replaced by some constant (independent of the
distance between ${\bf r}$ and ${\bf r'}$):
in this case they would be similar to a Hartree potential,
that exhibits the desired dependence upon surface charge.
A partly delocalized exchange-correlation hole would lead to the same kind of
behaviour~\cite{Resta96}.
But the pair-correlation for a homogeneous gas, $G^{\rm hom}$,
decays on average as the
{\it inverse of the fifth power} of
the distance between the points~\cite{WDA}.
This decay will make
these terms insensitive to the surface charge.

In the third term of Eq.~(\ref{Eq.WDA}),
the derivative of the pair-correlation function with respect
to some change of density appears. Ortiz and Martin have interpreted
the corresponding term in the exact expression as coming from the
polarizability of the exchange-correlation hole~\cite{Martin96},
sensitive to the TC or KS homogeneous electric field. But
we will see that the
WDA exchange-correlation hole is not as sensitive
as the exact exchange-correlation hole.
The derivative of the pair-correlation function can be obtained thanks to
the chain rule,
\begin{eqnarray}
 \frac{ \delta G^{\rm hom}({\bf r},{\bf r'};\bar n({\bf r};[n])) } { \delta
n({\bf r}_0) }
=
 \frac{ \delta G^{\rm hom}({\bf r},{\bf r'};\bar n({\bf r})) } { \delta \bar
n({\bf r}) }
\frac{ \delta \bar n({\bf r};[n])) } { \delta n({\bf r}_0) }.
\label{Eq.ChainRuleWDA}
\end{eqnarray}
The above-mentioned decay of $G^{\rm hom}$ affects both factors of the
right-hand side.
The first factor will exhibit the same decay as the homogeneous
gas pair-correlation function, while
the dependence of $\bar n({\bf r};[n])$
upon $\delta n({\bf r}_0)$, driven by the
requirement that the exchange-correlation hole integrates to $-1$, will also
be spatially short-ranged: a finite change of surface charge density on the
surface of the crystal,
times the $r^{-5}$ decay and integrated over the whole surface,
yields zero contribution in the thermodynamical limit.
As a consequence, all the terms in Eq.~(\ref{Eq.WDA})
are insensitive to arbitrarily remote surface charges, so that no
polarization-dependence is present in $V_{\rm xc}^{\rm WDA}$.

Thus, in opposition to the suggestion of
Mazin and Cohen~\cite{Mazin97,againstMazin}, we find that
the possible non-locality of approximate functionals is not sufficient to
generate
a polarization-dependence: one needs an ultra-non-local dependence.
This might be attained by considering either a model pair-correlation function
that does not integrate to $-1$ in the bulk~\cite{Resta96},
or a polarizable exchange-correlation hole~\cite{Martin96}, as in the
exact exchange approach~\cite{OEP}.

\section{The metal -- insulator paradox}

Godby and Needs~\cite{Godby89} observed that
the KS band structure of semiconductors might present the characteristics
of a metallic state (the absence of a band gap).
We now argue that it is precisely the {\it same} ${\cal O}(1/q^2)$
behaviour of $K_{\rm xc}(q,q)$ that allows
this ``true insulator--KS metal'' paradox to be understood.

In order to make the analysis as simple as possible,
we first ignore local field corrections, and also impose
cubic symmetry. In this case, only the head of the different
matrices appearing in Eq.~(\ref{X-Xo}) must be taken into account,
which means that this equation reduces to a scalar equality.
The small wavevector behaviours of $\chi$ and $\chi_o$ are well-known
in both the metallic and the insulating case~\cite{Pick70}.
The independent-particle polarizability of
a KS metallic ground-state
behaves like
\begin{equation}
\lim_{q \rightarrow 0} \chi_o(q,q)= \gamma,
\end{equation}
while for a cubic KS insulator,
one has
\begin{equation}
\lim_{q \rightarrow 0} \chi_o(q,q)= \alpha q^2,
\end{equation}
where $\gamma$ and $\alpha$ are some negative constants. The head of
the polarizability matrix $\chi(q,q)$ for a metal
(interacting electrons, not KS electrons)
behaves exactly as
$ -\frac{q^2}{4 \pi}$ in the long-wavelength limit (which corresponds to
complete screening of the Coulomb potential), while for cubic insulators,
it is $ -\frac{q^2}{4 \pi} \beta$, where $\beta$ is a positive
constant smaller than one,
describing the incomplete screening.

Now, one imposes
no divergence in $\chi_o^{-1}(q,q)$ (metallic KS ground state),
and non-cancelling divergences of $\chi^{-1}(q,q)$
and $V_{\rm C}$ (incomplete screening of the insulating system),
so that Eq.~(\ref{X-Xo}) without local fields becomes
\begin{equation}
\frac{1}{\gamma}= -\frac{4 \pi}{\beta q^2}+ \frac{4 \pi}{q^2} + K_{\rm
xc}(q,q),
\end{equation}
which proves that $K_{\rm xc}(q,q)$ {\it must} have
a ${\cal O}(1/q^2)$ divergence.
This result establish a connection between the ``true insulator--KS metal''
paradox
and the polarization dependence of the XC energy for insulators:
in both cases, the
small-wavevector behaviour of the XC kernel is similar.

\section{Unified treatment of $K_{\rm xc}$ for insulators and metals}

We now consider the treatment of insulators and metals
in a unique framework. By the way,
we will also generalize the demonstration contained in the
preceeding section to the case where local fields are included
(we treat not only the head
of $K_{\rm xc}$, $\chi_o^{-1}$
and $\pi^{-1}$, but also their wings and body).

Following the analysis by Pick, Martin and Cohen~\cite{Pick70}
of $\chi_o$ and $\pi$, we find that $\chi_o^{-1}$, for the KS
non-interacting system, and $\pi^{-1}$, for the interacting system, have
the following, similar,
non-analytic behaviour. In the insulating case, the head of these inverse
matrices
diverges like ${\cal O}(1/q^2)$, the wing elements diverge like ${\cal O}(1/q)$,
while the body elements, though non-analytic, are non-divergent. In the
metallic case, no
element shows a divergence for small wavevectors. The divergences in
$K_{\rm xc}=\chi_o^{-1} - \pi^{-1}$, following Eq.(\ref{Eq.11}), are easy
to deduce
from these results, if we suppose that there is no fortuitous cancellation
of divergences when the difference is taken.

For the (usual) ``true metal--KS metal'' case,
since there is no divergence in $\chi_o^{-1}$ or $\pi^{-1}$,
no element of $K_{\rm xc}$ need diverge for small wavevectors in order
to correct the long-wavelength response (although separate non-analytic
behaviour at non-zero wavevectors may be required to
accommodate a disparity between the
true and KS Fermi surfaces).
In all the other conceivable cases
(the usual ``true insulator--KS insulator'' case,
the exotic ``true insulator--KS metal'' case,
and the hypothetical ``true metal--KS insulator'' case),
the head of $K_{\rm xc}$ will
diverge like ${\cal O}(1/q^2)$, the wing elements will
diverge like ${\cal O}(1/q)$, and the body elements will not diverge.

Although the exchange-only kernel is able to generate the right divergences
in the ``true insulator--KS insulator'' case~\cite{OEP}, it is unable
to do so in
the ``true insulator--KS metal'' case. Indeed, the argument
developed in Ref.~\cite{OEP} relies on the following result~:
\begin{eqnarray}
V_{\rm x}({\bf q})=\sum_{\bf G} \chi_0^{-1}({\bf q},{\bf q}+{\bf G})   \; \;
     \frac{\delta E_{\rm x}}{\delta V_{\rm KS}({\bf q}+{\bf G})}
\end{eqnarray}
If the Kohn-Sham system is insulating, $\chi_o^{-1}$ will exhibit
the appropriate divergences. Following Ref.~\cite{OEP},
the exchange-hole polarizes in a homogeneous KS electric field,
so that one obtains a polarization-dependence of the exchange-only kernel.
On the other hand, if the Kohn-Sham system is metallic to start with,
$\chi_o^{-1}$ has no divergence, and
$V_{\rm x}({\bf q})$ for short wavevectors will vanish.

Thus, the corrections needed to obtain an insulating-like polarizability while
the Kohn-Sham system is metallic are entirely due to the correlation kernel.
In the adiabatic coupling constant construction of the XC energy~\cite{WDA},
the system at different value of the coupling constant $\lambda$ will
undergo a metal-insulator phase transition~: at $\lambda=0$, the characteristic
system is the Kohn-Sham metallic system, while at $\lambda=1$, one
deals with the truly interacting, insulating, system.

\section{ The macroscopic dielectric constant}

Now that the correct behaviour of $K_{\rm xc}$ has been
discussed, we would like to
investigate its consequences in the calculation
of the macroscopic dielectric constant of insulators.
At the macroscopic level, this quantity can be obtained as~\cite{Kittel68}:
\begin{equation}
\label{Eq.e-elec}
\varepsilon_{\infty} = 1 +
      4 \pi \frac{\partial {\cal P}}{\partial {\cal E}}
\end{equation}
where ${\cal P}$ is the macroscopic polarization and ${\cal E}$
is the macroscopic electric field.

In order to include the local-field effects in
the computation of $\varepsilon_{\infty}$, Adler and Wiser~\cite{Adler62+}
relied on the dielectric matrix, relating effective and external
potentials as follows
\begin{equation}
\label{Eq.e-1}
\sum_{\bf q+G'}
\varepsilon({\bf q+G, q+G'})
\delta V_{\rm eff}({\bf q+G'})
=
\delta V_{\rm ext} ({\bf q+G}),
\end{equation}
and connected the macroscopic dielectric constant
to the head of the inverse dielectric matrix:
\begin{equation}
\label{Eq.e-inf}
\varepsilon_{\infty} = \lim_{q \rightarrow 0}
\frac{1}{\varepsilon^{-1}({\bf q, q})}.
\end{equation}
The demonstration of Eq.~(\ref{Eq.e-inf})
was reported at the RPA level. The next step was
to include correctly the correction induced by the exchange-correlation
effects \cite{Singhal76}.

In the {\it test-charge} formulation of $\varepsilon$~\cite{Singhal76},
the effective potential appearing in Eq.~(\ref{Eq.e-1}) is chosen
as the one experienced by a hypothetical {\it
classical} charge, $\delta V_{\rm TC}=\delta V_{\rm ext}+\delta V_{\rm H}$.
Imposing the variations of the external potential
$\delta V_{\rm ext} ({\bf q+G})$ for ${\bf G} \neq {\bf 0}$ to be zero,
one deduces
\begin{eqnarray}
\frac{1}{\varepsilon^{-1}_{\rm TC}({\bf q, q})}
&=&
1- \frac{\delta V_{\rm H}({\bf q})}
        {\delta V_{\rm ext}({\bf q})+\delta V_{\rm H}({\bf q})}
\label{Eq.eTC}\\
&=&
1- \frac{4 \pi}{q^2}  \frac{\delta n({\bf q})}
                           {\delta V_{\rm ext}({\bf q})+\delta V_{\rm
H}({\bf q})}
\end{eqnarray}
The relationship that exists in the long-wave approach between
field and potential and between charge and polarization
allows one to recover Eq.~(\ref{Eq.e-elec}).

In the {\it electron} formulation, the effective potential appearing in
Eq.~(\ref{Eq.e-1}) is replaced by the one felt by the Kohn-Sham electrons,
$\delta V_{\rm KS}=\delta V_{\rm ext}+\delta V_{\rm H}+\delta
V_{\rm xc}$.
The dielectric constant is then~:
\begin{eqnarray}
\frac{1}{\varepsilon^{-1}_{e}({\bf q,q})}
&=&
1- \frac{\delta V_{\rm H}({\bf q})+\delta V_{\rm xc}({\bf q})}{\delta V_{\rm
ext}({\bf q})+\delta V_{\rm H}({\bf q})+\delta V_{\rm xc}({\bf q})}.
\label{Eq.e-e}
\end{eqnarray}

In the absence of polarization-dependence of $E_{\rm xc}$
($\delta V_{\rm xc}({\bf q})=0$), Eq.~(\ref{Eq.e-e}) reduces to
Eq.~(\ref{Eq.eTC}). Within the LDA, GGA or WDA,
{\it the test-charge and electron
definitions of $\varepsilon_{\infty}$ are therefore identical}.

Going beyond these approximations, we observe that the electron definition,
Eq.~(\ref{Eq.e-e}), does
not reduce to the test-charge one, Eq.~(\ref{Eq.eTC}), by a simple addition
of the
exchange-correlation
contribution in the denominator (macroscopic field),
but that the numerator must also include a modified Coulomb
interaction.
In addition, it can be checked that the test-charge definition of the
macroscopic
dielectric constant can be formulated as a second derivative of the
(electric) free
energy~\cite{deGironcoli89} with respect to a homogeneous electric field, which
is not the case for the electron formulation.

It is also possible to analyze the direct effect
of the divergence of head and wing elements of $K_{\rm xc}(q,q)$ on dielectric
matrices. Following Singhal and Callaway~\cite{Singhal76}, we use the equality
\begin{equation}
\delta n
= \chi_o [\delta V_{\rm ext}+\delta V_{\rm H}+\delta V_{\rm xc}]
\end{equation}
to find the form of
the dielectric matrices in terms of $\chi_o$, $V_C$ and $K_{\rm xc}$~:
\begin{equation}
\varepsilon_{\rm TC} = 1- V_C \chi_o \; [1- K_{\rm xc} \chi_o]^{-1},
\label{Eq.eTC2}
\end{equation}
\begin{equation}
\varepsilon_{e} =  1- V_C \chi_o - K_{\rm xc} \chi_o  .
\label{Eq.eel}
\end{equation}
These expressions are governed by the products
$V_C \chi_o$ and $K_{\rm xc} \chi_o$.
The head and wing elements of $\chi_o$ behave
like ${\cal O}(q^2)$ and ${\cal O}(q)$ in the
limit of $q \rightarrow 0$~\cite{Pick70,Hybertsen87}.
$V_C$ is diagonal, with a divergent ${\cal O}(1/q^2)$ head.
In the small wavevector limit, their product, an asymmetric matrix,
will have finite head and body, while the upper wing will diverge
like ${\cal O}(1/q)$ and the lower wing will behave like ${\cal O}(q)$.
If we completely ignore the effect of the exchange-correlation
terms in Eqs.~(\ref{Eq.eTC2}) and~(\ref{Eq.eel}),
the latter form is also the form of the dielectric matrices.

If we take into account the effect of a {\it non-divergent} $K_{\rm xc}$,
of the LDA type, the product $K_{\rm xc} \chi_o$ will behave
like $\chi_o$. The handling of such contributions
in both Eqs.~(\ref{Eq.eTC2}) and~(\ref{Eq.eel}) will only
affect the body of the dielectric matrices, in the small wavevector limit.
By contrast, if we take into account the possible divergences of $K_{\rm xc}$,
the product $K_{\rm xc} \chi_o$ will behave like $V_C \chi_o$.
As such, it will be able to affect the leading behaviour of all elements in
both formulations of the dielectric matrix.

The macroscopic dielectric constant being obtained
from the head of the {\it inverse} dielectric matrix Eq.~(\ref{Eq.e-inf}), the
presence of $K_{\rm xc}$ will always affect its amplitude. However,
if one deals with a LDA-type $K_{\rm xc}$,
its influence will be mediated by local fields only ,
while the true $K_{\rm xc}$ will
modify it more directly, in particular through a modification of the head
of the
dielectric matrix.

\section{Conclusions}

In conclusion, we have seen that the polarization dependence of  $E_{\rm xc}$
imposes a condition on the form of the exchange-correlation kernel: its head
must diverge like ${\cal O}(1/q^2)$ in the limit of $q \rightarrow 0$. This
condition is not satisfied within the LDA, GGA and WDA.
The absence of this divergence in the approximate functionals will
affect the amplitude of the computed macroscopic dielectric constant by a
finite amount. A similar divergence is also needed, in order to reproduce the
correct dielectric response, when insulators are described as metals in
Kohn-Sham
theory.

\acknowledgments
We acknowledge numerous discussions with R. Martin,
D. Vanderbilt, and R. Resta.
We thank the latter for a careful reading of the final manuscript,
Yong-Hoon Kim for questioning the effect of the
exchange-only kernel in the metallic case,
and the authors of Refs.~\cite{Aulbur97,Resta96,Martin96,Martin97}
for providing copies of their
manuscripts prior to publication. We acknowledge financial support from
FNRS-Belgium and the Concerted Action programme 92-97.156 (X.G.), the
EPSRC (R.W.G.), the European Union
(HCM Program Contract CHRX-CT940462), and the
Academic Research Collaboration Programme between the FNRS, the CGRI, and
the British Council.

\end{multicols}

\begin{figure}
\caption{The diagonal part of the exchange-correlation kernel for the
one-dimensional model semiconductor. The result was obtained for a
80 unit-cell supercell (320 a.u.). The ${\cal O}(1/q^2)$ character of the
$K_{\rm xc}$ divergence is exhibited in the inset by the non-zero intercept of
$q^2  .  K_{\rm xc}(q,q)$.}
\label{Fig.Kxc}
\end{figure}


\begin{thebibliography}{10}
\bibitem{sign}
Note that $n$ is an electronic density, not a charge
density. This equation differs from Eq.(3) in Ref.~\cite{Gonze95}
because the choice of sign
for the definition of the polarization in the latter paper
was unconventional, albeit treated in a consistent way.
\bibitem{Comment2}
In Ref.~\cite{Mazin97}, it was pointed out that the
``exchange-correlation local field'' (XCLF) correction to the
Green's function approach diverges like ${\cal O} (1/q^2)$.
Although this additional effect is not present
in our model, a ${\cal O} (1/q^2)$ divergence is still observed.
\bibitem{againstMazin}
We disagree with other statements in Ref.~\cite{Mazin97}.
For example, there are semiconductors described as metals in LDA,
with formally infinite LDA dielectric constants, in contrast
with the supposedly excellent accuracy of LDA mentioned in
the beginning of Ref.~\cite{Mazin97}. This problem, in the context of
exact KS theory,
is analyzed in the present paper. Also, it is claimed in Ref.~\cite{Mazin97}
that the most divergent term in the XC kernel is a delta function. Actually,
in that paper,
it is only proved that the XC kernel is infinite at $q=0$, so that
nothing prevents the stronger ${\cal O}(1/q^2)$ behaviour
observed here.
Finally, Ref.~\cite{Mazin97} fails to distinguish clearly between the limit
$q \rightarrow 0$ and the case $q=0$, particularly when attributing
to Ref.~\cite{Gonze95} results about
the validity of conventional DFT.  We are in agreement with the alternative
discussion
of these issues, and their relationship to Ref.~\cite{Gonze95}, given by Martin
and Ortiz~\cite{Martin97}.
\end{thebibliography}
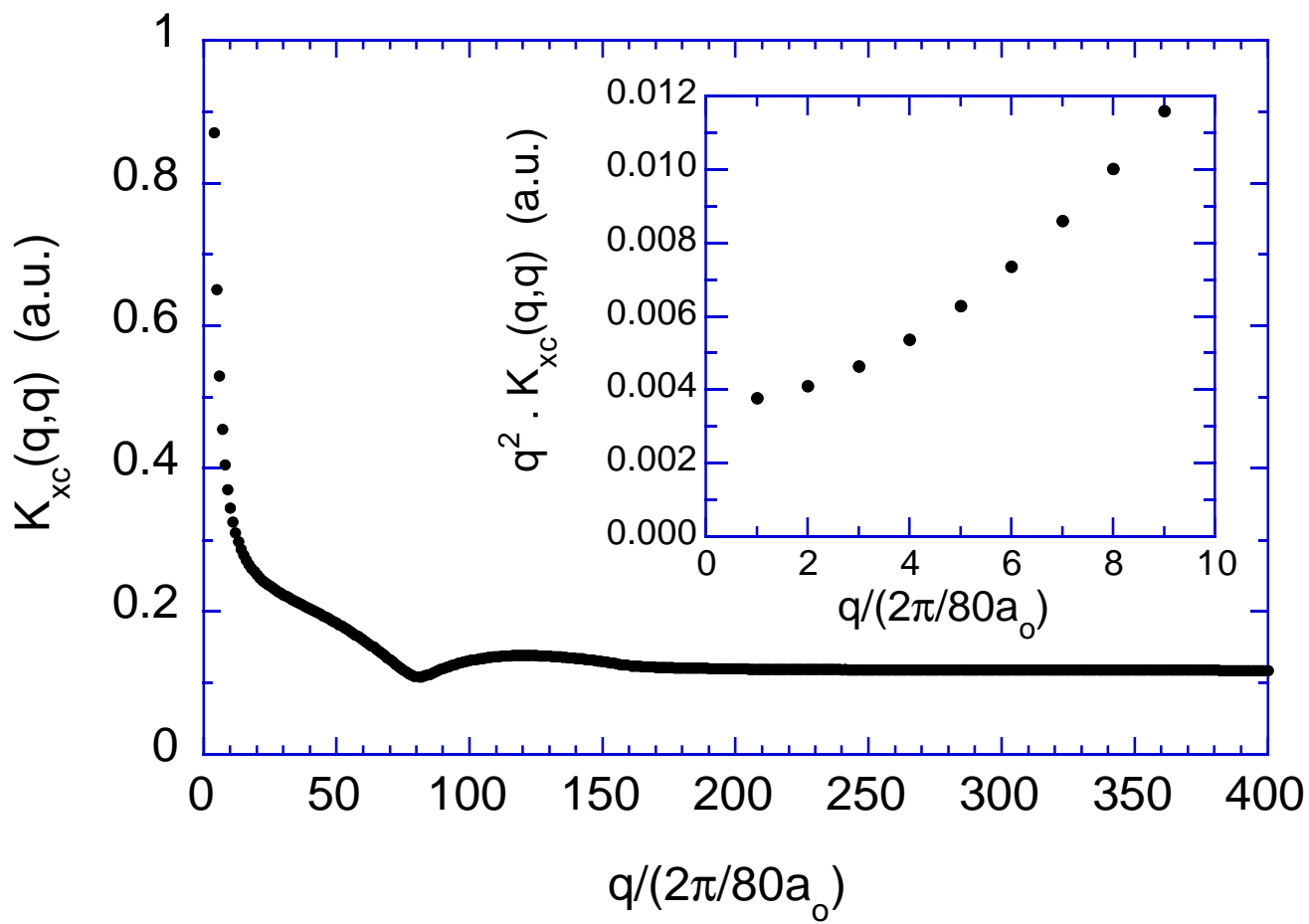
\end{document}